\documentclass[aps,prd,twocolumn,groupedaddress,showpacs]{revtex4-1}
\usepackage{epsf,epsfig,graphicx}
\usepackage{amsmath}
\usepackage{amssymb}
\usepackage{tikz}

\def\edth{\;\raise1.0pt\hbox{$'$}\hskip-6pt\partial\;}
\def\baredth{\;\overline{\raise1.0pt\hbox{$'$}\hskip-6pt
\partial}\;}
\def\gsim{~\rlap{$>$}{\lower 1.0ex\hbox{$\sim$}}}

\newcommand{\be}{\begin{equation}}
\newcommand{\ee}{\end{equation}}
\newcommand{\bw}{\begin{widetext}}
\newcommand{\ew}{\end{widetext}}

\usepackage{xcolor}

\usepackage[colorlinks]{hyperref}
\definecolor{darkblue}{HTML}{2E3092}
\hypersetup{colorlinks=true,linkcolor=darkblue,citecolor=darkblue,urlcolor=darkblue}

\AtBeginDocument{}

\begin{document}

\title{Redshift-space fluctuations in stochastic gravitational wave background}

\author{Kin-Wang Ng$^{1,2}$}

\affiliation{
$^1$Institute of Physics, Academia Sinica, Taipei 11529, Taiwan\\
$^2$Institute of Astronomy and Astrophysics, Academia Sinica, Taipei 11529, Taiwan
}

\vspace*{0.6 cm}
\date{\today}
\vspace*{1.2 cm}

\begin{abstract}
We study the redshift-space fluctuations induced by a stochastic gravitational wave background (SGWB) 
via the Sachs-Wolfe effect.
The redshift-space fluctuations can be encapsulated in a line-of-sight integral that is useful for studying the imprint of short-wavelength gravitational waves on the cosmic microwave background (CMB) anisotropy.
We thus derive constraints on the SGWB from small-scale CMB anisotropy measurements.
Our results reproduce the constraint on the short-wavelength SGWB, previously derived from the Planck and BICEP/Keck array CMB data with a CMB Boltzmann numerical code. 
Furthermore, we improve the constraint and extend it to shorter wavelengths by using the CMB measurements made by the Atacama Cosmology Telescope and the South Pole Telescope.
Also, the integral provides us with a precise redshift fluctuation correlation between a pair of pulsars in pulsar timing measurements, which conveniently incorporates the effect of the pulsar term into a small-angle correlation.
We further discuss the observation of pulsar pairs in globular clusters to look for this small-angle correlation.
\end{abstract}

\maketitle

\section{Introduction}

The search for stochastic gravitational wave background (SGWB) is one of the main goals in observational cosmology. After the discovery of GWs emitted by a binary black hole merger made by the LIGO-Virgo Collaboration~\cite{ligo} and the observation of a handful of GW events from compact binary coalescences~\cite{ligo2019}, the detection of the SGWB becomes the next milestone in a new era of GW astronomy and cosmology.
There have been many studies on possible astrophysical and cosmological sources for the SGWB such as distant compact binary coalescences, early-time phase transitions, cosmic string or defect networks, second-order primordial scalar perturbations, and inflationary GWs~\cite{romano}. GWs have very weak gravitational interaction, so they decouple from matter at the time of production and travel to us almost without being disturbed. At present, they remain as a GW background that encodes the information of the production processes in the early Universe. 

The spectrum of the SGWB is expected to span a wide range of frequencies. The method adopted in the GW interferometry such as the LIGO-Virgo experiment for detecting the SGWB is to correlate the responses of a pair of detectors to the GW strain amplitude. The correlation allows us to filter out detector noises and obtain a large signal-to-noise ratio for the detection of GWs of frequencies at several tens hertz~\cite{romano}. An indirect method to search for the SGWB is through their gravitational effects on physical observables such as the cosmic microwave background (CMB)~\cite{kamion} and the arrival times of radio pulses from millisecond pulsars~\cite{romano}. Horizon-sized GWs can leave an imprint on the anisotropy and polarization of the CMB that has been long sought after in CMB experiments, whereas the pulsar timing is sensitive to short-wavelength GWs at nanohertz frequencies. 
Future GW experimental plans such as Einstein Telescope, Cosmic Explorer, LISA, DECIGO, Taiji, TianQin, international pulsar-timing arrays, and SKA~\cite{ligo2050}, hand in hand with CMB Stage-4 experiments~\cite{cmb4}, will certainly bring us a precision science in SGWB observation.

In this paper, we will give a systematic study of the gravitational effects induced by the SGWB on astrophysical and cosmological observables. The study will be directly applied to the indirect measurements of the SWGB in CMB small-scale anisotropy experiments and in pulsar-timing-array observation. Constraints on the SWGB from CMB data have been extensively studied mostly using CMB numerical Boltzmann codes~\cite{lasky16,planck18_sgwb,namikawa}; however, difficulties arise in short-wavelength regimes due to heavy cancellations in mode projection~\cite{namikawa}. Therefore, we give up on this, rather relying on a single line-of-sight integral to compute CMB anisotropy power spectra induced by short-wavelength SGWB. We will see that this analytic approach reproduces the results of 
Ref.~\cite{namikawa} and enables us to extend the CMB constraints to a very short-wavelength SGWB. Furthermore, the line-of-sight integral is in fact the integrated form of the Shapiro time delay of the arrival times of radio pulses from pulsars. It is known that the earth term in the Shapiro time delay leads to the Hellings and Downs curve for the interpulsar correlation~\cite{downs}, while the pulsar term adds power to the correlation at small separation angles~\cite{mingar14,chu2107}. However, the effect of the pulsar term in terms of power spectrum has been scarcely studied. 
We will find that the line-of-sight integral can conveniently incorporate the effect of the pulsar term into the interpulsar correlation. It can reproduce the power spectrum of the Hellings and Downs curve on large angular scales found in Ref.~\cite{gair} and add power to the power spectrum at small-scales induced by the pulsar term.

In the next section, we firstly review the propagation of free GWs in the expanding universe. In Sec.~\ref{RSF}, the effect on the redshift space due to the presence of a SGWB is discussed. Then, this is applied to the induced CMB anisotropy in Sec.~\ref{cmba} and pulsar timing in Sec.~\ref{ptiming}.
Section~\ref{conclusion} is our conclusion.

\section{Stochastic Gravitational Wave Background}

Consider a perturbed metric:
\begin{equation}
ds^2=-a^2 d\eta^2+ a^2(\delta_{ij}+h_{ij})dx^idx^j,
\end{equation}
where $a(\eta)$ is the cosmic scale factor and $\eta$ is the conformal time defined by $d\eta=dt/a$. The transverse-traceless tensor perturbation $h_{ij}$ can be decomposed into two independent polarization tensors as
\begin{equation}
h_{ij}(\eta,\vec{x})=\sum_\lambda \int\frac{d^3\vec{k}}{(2\pi)^{3\over2}} \left[ a_\lambda(\vec{k}) h_\lambda(\eta,\vec{k}) \epsilon_{ij}^\lambda (\hat{k}) e^{i\vec{k}\cdot\vec{x}} + H.c.\right],
\label{hij}
\end{equation}
where $\epsilon_{ij}^\lambda (\hat{k})\epsilon_{ij}^{\lambda'}(\hat{k})=2\delta_{\lambda\lambda'}$. The annihilation and creation operators,
$a_\lambda(\vec{k})$ and $a^\dagger_\lambda(\vec{k})$ respectively, satisfy the commutation relation,
\begin{equation}
\left[a_\lambda(\vec{k}), a^\dagger_{\lambda'}(\vec{k}')\right]= \delta(\vec{k}-\vec{k}')  \delta_{\lambda\lambda'}.
\end{equation}
The GW amplitude, $h_\lambda(\eta,\vec{k})$, is governed by the equation of motion,
\begin{equation}
\frac{d^2 h_\lambda}{d\eta^2} +\frac{2}{a} \frac{da}{d\eta} \frac{d h_\lambda}{d\eta} +k^2 h_\lambda =0.
\label{EOM}
\end{equation}
The spectral energy density of the SGWB relative to the critical density is then given by
\begin{eqnarray}
\Omega_{\rm GW} (\eta,k,\hat{k}) &\equiv& \frac{k}{\rho_c} \frac{d\rho_{\rm GW}}{dkd^2\hat{k}}
=\sum_\lambda \frac{1}{12a^2 H^2} \left(\frac{k}{2\pi}\right)^3 \nonumber \\
&&\times \left[k^2|h_\lambda|^2+\left\vert\frac {d h_\lambda}{d \eta}\right\vert^2 \right],
\end{eqnarray}
where $\rho_c=3M_p^2H^2$, with $M_p$ being the reduced Planck mass. Writing $h_\lambda=k^{-3/2}h$, we have
\begin{equation}
\Omega_{\rm GW} (\eta,k,\hat{k}) = \frac{1}{48\pi^3} \left({k\over aH}\right)^2 
\left[|h|^2+\left\vert\frac{1}{k}\frac {d h}{d \eta}\right\vert^2 \right],
\label{omegaGW}
\end{equation}
and the tensor power spectrum is defined as ${\cal P}(\eta,k)\equiv |h(\eta,\vec{k})|^2/(2\pi^2)$. 
The $h(\eta,\vec{k})$ is dispersive and it can be cast into $h(\eta,\vec{k})=h(k\eta)$. 
For a superhorizon mode with $k\eta\ll 1$, $h(k\eta)$ has a constant amplitude; 
$h(k\eta)$ then oscillates with a decaying envelope once the mode enter the horizon.
For example, in slow-roll inflation models, metric quantum fluctuations during inflation give rise to
an initial condition of the GW amplitude for superhorizon modes:
\begin{equation}
\vert h(k\eta)\vert = \frac{H_I}{M_p}\quad{\rm for}\quad{k\eta\ll 1},
\end{equation}
where $H_I$ is the Hubble scale in inflation. This implies a scale-invariant power spectrum,
\begin{equation}
 {\cal P}(k)\equiv {\cal P}(\eta,k)\big\vert_{k\eta\ll 1} = \frac{1}{2\pi^2}\frac{H_I^2}{M_p^2}.
\label{SIpower}
\end{equation}
Another kind of the SGWB may be generated in a physical process taking place within the horizon with a characteristic frequency $k_*$ at time $\eta_*$:
\begin{equation}
\vert h(k_*\eta_*)\vert = \frac{M_*}{M_p} \quad{\rm with}\quad{k_*\eta_* > 1},
\end{equation}
where $M_*$ represents some mass
 scale. This results in a narrow initial power spectrum with a peak height:
\begin{equation}
{\cal P}(\eta_*,k_*) = \frac{1}{2\pi^2}\frac{M_*^2}{M_p^2}.
\label{Npower}
\end{equation}
The subsequent time evolution of $h(k_*\eta)$ is then determined by Eq.~(\ref{EOM}) for $\eta>\eta_*$. 
The solution for this subhorizon mode can be approximated as
\begin{equation}
h(k_*\eta)\simeq  \frac{M_*}{M_p} \frac{a(\eta_*)}{a(\eta)} e^{-ik_*\eta}.
\label{submode}
\end{equation}
From Eq.~(\ref{omegaGW}), the present spectral energy density for an isotropic SGWB is
\begin{equation}
\Omega_{\rm GW} \simeq  \frac{1}{6\pi^2} \frac{M_*^2}{M_p^2} \left[\frac{k_* a(\eta_*)}{k_0 a(\eta_0)}\right]^2,
\label{omegaGW2}
\end{equation}
where $k_0=a(\eta_0) H_0$ is the wavenumber of the mode that just crosses the present horizon.

\section{Redshift-space Fluctuations}
\label{RSF}

The gravitational effects due to the presence of a SGWB can be encoded in a fluctuation in the redshift of an observed photon source. Suppose the photon source locate at redshift $z$. Then, the fluctuation in the redshift of the photon source is given by the Sachs-Wolfe effect~\cite{sachs},
\begin{equation}
z+1=\frac{a(\eta_r)}{a(\eta_e)}\left[1-{1\over 2}\int_{\eta_e}^{\eta_r}d\eta\, e^ie^j
\frac{\partial}{\partial\eta}h_{ij}(\eta, \vec x) \right],
\end{equation}
where ${\bf e}$ is the propagation direction of the photon. The lower (upper) limit of integration in the line-of-sight integral represents the point of emission (reception) of the photon. Let ${\bar z}$ be the mean redshift and $\delta z=z-{\bar z}$ be the fluctuation. Then, we have $1+{\bar z}={a(\eta_r)}/{a(\eta_e)}$ and
\begin{equation}
{\Delta z}({\bf e})\equiv \frac{\delta z}{1+{\bar z}}({\bf e})= - {1\over 2}\int_{\eta_e}^{\eta_r}d\eta\, e^ie^j\frac{\partial}{\partial\eta}h_{ij}(\eta, \vec x).
\label{Deltaz}
\end{equation}
This redshift-space fluctuation can be expanded in terms of spherical harmonics,
\begin{equation}
{\Delta z}({\bf e}) = \sum_{l,m} a_{lm} Y_{lm}({\bf e}).
\end{equation}
For an isotropic unpolarized SGWB, the isotropy in the mean guarantees that
\begin{equation}
\langle a^{\dagger}_{lm} a_{l'm'}\rangle=C_l \delta_{ll'}\delta_{mm'},
\end{equation}
where $C_l$ is the redshift-space anisotropy power spectrum, from which we
can construct the two-point correlation function,
\begin{equation}
 \langle {\Delta z}({\bf e_1}) {\Delta z}({\bf e_2}) \rangle=\sum_l \frac{2l+1}{4\pi} C_l P_l({\bf e}_1\cdot{\bf e}_2),
 \label{corr}
\end{equation}
where $P_l$ is the Legendre polynomial.  Using Eq.~(\ref{hij}) and doing the tensor contraction, 
we obtain the formula for the power spectrum as~\cite{abbott}
\begin{eqnarray}
C_l&=&\frac{1}{2\pi} (l+2)(l+1)l(l-1)\times \nonumber \\
&&\int_0^\infty \frac{dk}{k}
    \left\vert\int_{\eta_e}^{\eta_r} d\eta \frac{dh(k\eta)}{d\eta}
    \frac{j_l[k(\eta_r-\eta)]}{k^2(\eta_r-\eta)^2}\right\vert^2,
\label{cmbCl}
\end{eqnarray}
where $j_l$ is a spherical Bessel function. 

\section{CMB temperature anisotropy}
\label{cmba}

The redshift-space fluctuations can induce a temperature anisotropy of the CMB, given by Eq.~(\ref{Deltaz})
\begin{equation}
\frac{\delta T}{T}({\bf e})={\Delta z}({\bf e}),
\end{equation}
where $\eta_e=\eta_{\rm dec}$ denoting the CMB decoupling time and $\eta_r=\eta_0$ the present time. This is the well-known Sachs-Wolfe tensor contribution to the CMB temperature anisotropy, whose power spectrum is then given by Eq.~(\ref{cmbCl}).

\subsection{A scale-invariant power spectrum}

The CMB temperature anisotropy due to the primordial tensor power spectrum~(\ref{SIpower}) 
has been well studied (see, for example, Ref.~\cite{abbott}). 
Here we recapitulate the main results for completeness. Also, they serve the purpose of defining the time and length scales used below and are useful for us to understand the discussions later on. 
For a fixed $l$, the main contribution to the integral~(\ref{cmbCl}) for $C_l$ comes from the mode of wavenumber 
$k\simeq l/\eta_0$ at the horizon crossing time $\eta_c\simeq \pi/k$ \cite{star2}. 
Since a mode is dispersive after entering the horizon, 
the modes that can imprint a large anisotropy on the CMB should have $\eta_c > \eta_{\rm dec}$. 
In the standard $\Lambda$CDM model~\cite{planck18}, $\eta_{\rm dec}\simeq 300\,{\rm Mpc}$ and 
the comoving distance to the CMB decoupling surface is $\eta_0-\eta_{\rm dec}\simeq \eta_0\simeq 14000\,{\rm Mpc}$,
where we have chosen $a(\eta_0)=1$.
This explains why superhorizon modes with $k<\pi/\eta_{\rm dec}\simeq 0.01\,{\rm Mpc}^{-1}$ dominate the contribution to the CMB temperature anisotropy on large angular scales at $l< \pi\eta_0/\eta_{\rm dec} \simeq 150$. 

\subsection{A narrow power spectrum}

For the narrow power spectrum~(\ref{Npower}), we adopt the subhorizon-mode solution~(\ref{submode}).
The induced CMB anisotropy power spectrum is then given by
\begin{eqnarray}
C_l&\simeq&\frac{1}{2\pi} (l+2)(l+1)l(l-1)\times \nonumber \\
&&\Delta\ln{k_*}
    \left\vert\int_{\eta_1}^{\eta_0} d\eta \frac{dh(k_*\eta)}{d\eta}
    \frac{j_l[k_*(\eta_0-\eta)]}{k_*^2(\eta_0-\eta)^2}\right\vert^2,
\end{eqnarray}
where $\eta_1={\rm max}(\eta_{\rm dec},\eta_*)$. Assuming that the spectrum spans a range of $\Delta\ln{k_*}\simeq 1$ 
and that the universe was in a matter-dominated epoch with $a(\eta)=(\eta/\eta_0)^2$, we obtain
\begin{eqnarray}
C_l&\simeq&\frac{M_*^2}{2\pi M_p^2} \left(\frac{\eta_*}{\eta_0}\right)^4  (l+2)(l+1)l(l-1)\times \nonumber \\
&& \left\vert\int_{0}^{x_1} dx \frac{1-x/x_0-i2/x_0}{(1-x/x_0)^3} e^{ix} \frac{j_l(x)}{x^2}\right\vert^2,
\label{Cl4narrow}
\end{eqnarray}
where $x=k_*(\eta_0-\eta)$, $x_0=k_*\eta_0$, and $x_1=k_*(\eta_0-\eta_1)$.
We compute the power spectrum in two limiting cases:

\subsubsection{Pre-recombination with $\eta_*<\eta_{\rm dec}$}

\begin{figure}[htp]
\centering
\includegraphics[width=0.5\textwidth]{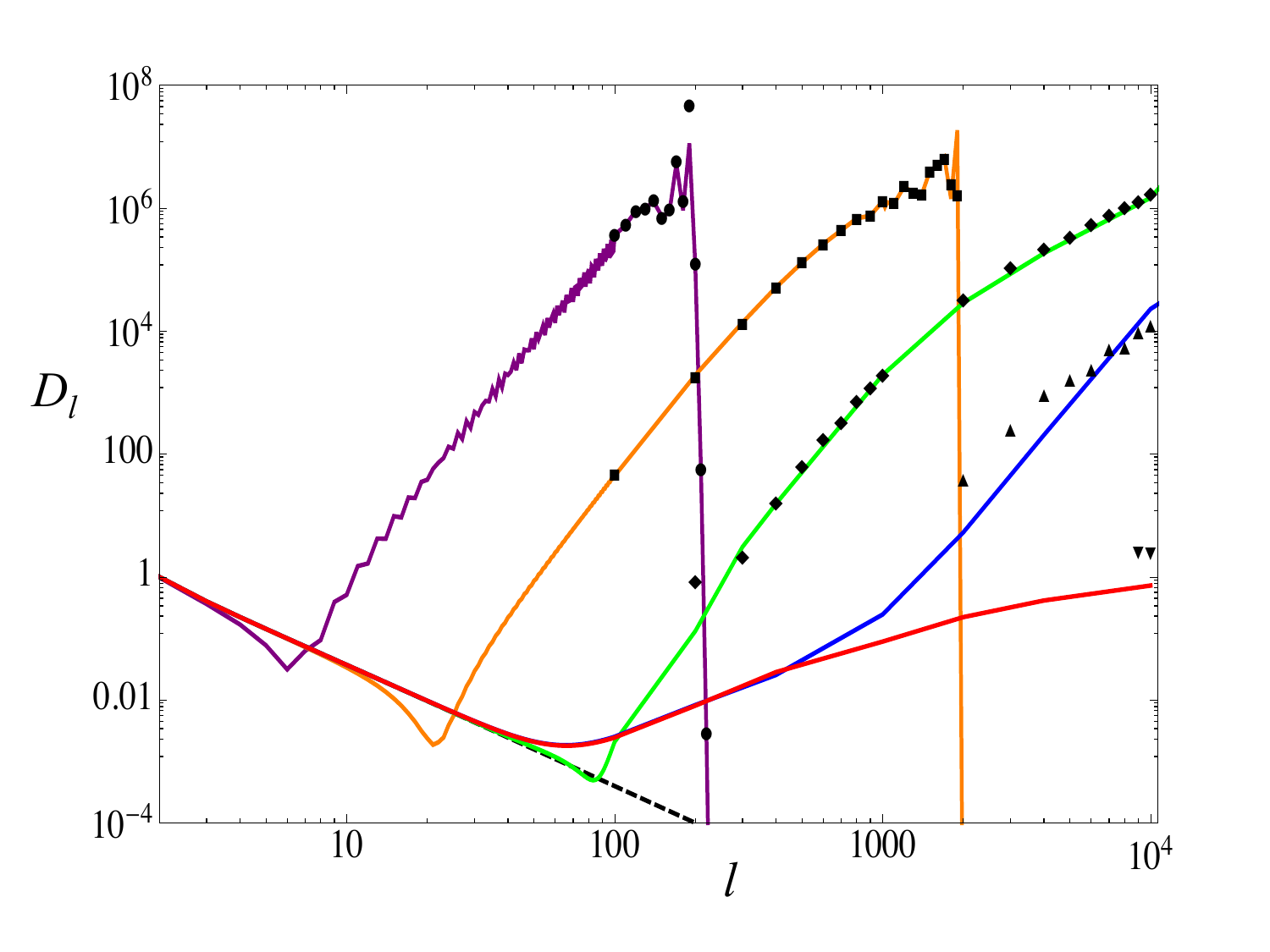}
\caption{CMB temperature anisotropy power spectra induced by narrow spectra of gravitational waves centered at wavelengths of $k_*=0.014$, $0.14$, $1.4$, $14$, and $140\,{\rm Mpc}^{-1}$, 
denoted by five solid curves from left to right, respectively. We have defined 
$D_l\equiv l(l+1)C_l (2\pi M_p^2 \eta_0^4)/(M_*^2 \eta_*^4)$, where $C_l$ is given by Eq.~(\ref{Cl4narrow}).
The plot markers near or at each solid curve are computed using the large-order approximation in Eqs.~(\ref{large-order1}), (\ref{large-order2}), and (\ref{large-order3}). 
The dashed line shows the $l^{-2}$ scaling on large angular scales.}
\label{fig1}
\end{figure}

In this case, we have $x_1/x_0=(\eta_0-\eta_{\rm dec})/\eta_0=137/140$. 
The integrand in Eq.~(\ref{Cl4narrow}) is highly oscillating functions which render heavy cancellation to the integration. Indeed, this makes a brute-force numerical integration very difficult especially for a large $k_*$. 
However, for a fixed $l$ the integral receives contributions only when $x\sim l$ and $x\sim x_1$. 
Thus, we have evaluated the integration over small ranges of $x$ covering the contributing regions and then increased the ranges to obtain values within the required accuracy. Using this strategy, we have computed the $C_l$ for 
$k_*=0.014$, $0.14$, $1.4$, $14$, and $140\,{\rm Mpc}^{-1}$, as shown in Fig.~\ref{fig1}. 
To assure the results for high-$l$ multipoles, we let $x=(l+1/2)y$ and approximate Eq.~(\ref{Cl4narrow}) as
\begin{eqnarray}
C_l&&\simeq\frac{M_*^2}{4M_p^2}\left(\frac{\eta_*}{\eta_0}\right)^4 (l+2)(l+1)l(l-1)(l+1/2)^2\times \nonumber \\
&& \left\vert\int_{0}^{y_1} dy \frac{1-x/x_0-i2/x_0}{(1-x/x_0)^3} e^{ix} 
\frac{J_{l+{1\over2}}[(l+{1\over2})y]}{x^{5/2}}\right\vert^2,
\label{large-order1}
\end{eqnarray}
where $J_\nu(\nu y)$ takes the asymptotic form for a large order as~\cite{Jfunction}
\begin{eqnarray}
J_\nu(\nu y)&\sim&\frac{e^{\nu(1-y^2)^{1\over2}-\nu\tanh^{-1}(1-y^2)^{1\over2}}}{\sqrt{2\pi\nu}\,(1-y^2)^{1\over4}}
\nonumber \\
&&{\rm for}\quad 0<y<1,\label{large-order2}\\
J_\nu(\nu y)&\sim& \frac{\cos[\nu(y^2-1)^{1\over2}-\nu\tan^{-1}(y^2-1)^{1\over2}-{\pi/4}]}{\sqrt{\pi\nu/2}\,(y^2-1)^{1\over4}} 
\nonumber \\
&&{\rm for}\quad y>1. \label{large-order3}
\end{eqnarray}
We have used this approximation to compute $C_l$'s, 
which are denoted by the plot markers near or at each solid curve in Fig.~\ref{fig1}. 
For $k_*=0.014$ and $0.14\,{\rm Mpc}^{-1}$, the approximation works very well. For $k_*=1.4\,{\rm Mpc}^{-1}$, it works well too except when $l\lesssim 200$.
For $k_*=14$ and $140\,{\rm Mpc}^{-1}$, it reproduces fairly well the $C_l$'s for $l=10^4$, while overestimating the relatively low-$l$ multipoles.

In Ref.~\cite{namikawa}, the authors have used the CAMB numerical code to compute the CMB anisotropy and polarization power spectra induced by a monochromatic SGWB produced before the time of decoupling. They have produced the $C_l$ for $l<2000$ at $k_*=0.014$, $0.14$, $1.4$, and $7.81\,{\rm Mpc}^{-1}$. 
The power spectra in Fig.~\ref{fig1} match fairly well with their results whenever the input parameters overlap. Here we have extended the range of the power spectra to $l\le10^4$ and $k_*\le 140\,{\rm Mpc}^{-1}$. For example, for $k_*=0.14\,{\rm Mpc}^{-1}$ the power spectrum $l(l+1)C_l$ peaks around $l\sim x_0=1960$, as expected for the short-wavelength modes that mainly contribute to the small-scale anisotropy. 
These short-wavelength modes can also contribute to the large-scale CMB anisotropy, resulting in a local maximum at $l=2$ and a local minimum at $l=21$, when the CMB photons arrive at the observer at the present epoch. This can be seen by taking the limit, $j_l(x)/x^2 \rightarrow x^{l-2}$ as $x\rightarrow 0$, in Eq.~(\ref{Cl4narrow}). We will further study this large-scale contribution in the next case.

\subsubsection{Post-recombination with $\eta_{\rm dec}<\eta_*\lesssim \eta_0$}

In this case, $x_1\ll x_0$ and the power spectrum can be approximated as
\begin{eqnarray}
C_l&\simeq&\frac{M_*^2}{2\pi M_p^2} \left(\frac{\eta_*}{\eta_0}\right)^4 (l+2)(l+1)l(l-1)\times \nonumber \\
&& \left\vert\int_{0}^{x_1} dx\, \left(1+\frac{2x}{x_0}+\frac{3x^2}{x_0^2}+..\right) e^{ix} \frac{j_l(x)}{x^2}\right\vert^2,
\label{Clexpand}
\end{eqnarray}
When $x_1\ll 1$, we have
\begin{equation}
C_l\simeq\frac{M_*^2}{2\pi M_p^2} \left(\frac{\eta_*}{\eta_0}\right)^4 
\frac{(l+2)(l+1)l(l-1)}{(2l+1)!!\,(2l+1)!!} x_1^{2l-2}.
\end{equation}
When $x_1\gg 1$, using the integral result,
\begin{eqnarray}
\int_{0}^{\infty} dx\, &&e^{-\alpha x} J_\nu(\beta x) x^{\mu-1} =
\frac{(\beta/2)^\nu}{\alpha^{\nu+\mu}} \frac{\Gamma(\nu+\mu)}{ \Gamma(\nu+1)} \times \nonumber \\
&&F\left(\frac{\nu+\mu}{2},\frac{\nu+\mu+1}{2};\nu+1;-\frac{\beta^2}{\alpha^2}\right),
\end{eqnarray}
where $F(a,b;c;d)$ is a hypergeometric function which has a particular value,
\begin{equation}
F(a,b;c;1)=\frac{\Gamma(c)\Gamma(c-a-b)}{\Gamma(c-a)\Gamma(c-b)},
\end{equation}
and the doubling formula for gamma functions,
\begin{equation}
\Gamma(2z)=\frac{2^{2z-1}}{\sqrt{\pi}} \Gamma(z) \Gamma(z+1/2),
\end{equation}
we obtain
\begin{eqnarray}
&&\int_{0}^{x_1} dx\, e^{ix} j_l(x) x^{\mu-{1\over2}} \nonumber \\
&\simeq& \sqrt{\pi\over2}\int_{0}^{\infty} dx\,e^{ix} J_{l+{1\over2}}(x) x^{\mu-1} \nonumber \\
&=&\frac{i^{l+\mu+{1\over2}}}{2^{\mu+{1\over2}}} \Gamma({1/2}-\mu)
\frac{\Gamma(l+\mu+{1/2})}{\Gamma(l-\mu+{3/2})},
\label{besselintegral}
\end{eqnarray}
where $\alpha=-i$, $\beta=1$, $\nu=l+{1/2}$, and we have approximated $x_1$ by an infinity.   
Under this approximation, we keep only the first and the second terms in Eq.~(\ref{Clexpand}) 
that correspond to $\mu=-{3/2}$ and $-{1/2}$, respectively. Hence, we have
\begin{eqnarray}
C_l\simeq\frac{2M_*^2}{\pi M_p^2} \left(\frac{\eta_*}{\eta_0}\right)^4 
\left[\frac{1}{(l+2)(l+1)l(l-1)}+\right.&& \nonumber \\
\left.\frac{1}{x_0^2}\frac{(l+2)(l-1)}{(l+1)l}\right].&&
\label{extraCl}
\end{eqnarray}
For $k_*>0.014\,{\rm Mpc}^{-1}$ that we consider here, $x_0=k_*\eta_0>196$. Thus, the first term dominates and $l(l+1)C_l$ scales as $l^{-2}$ for $l\lesssim \sqrt{x_0}$. This explains the large-scale power and scaling of the power spectra as shown by the dashed line in Fig.~\ref{fig1}.

\begin{figure}[htp]
\centering
\includegraphics[width=0.5\textwidth]{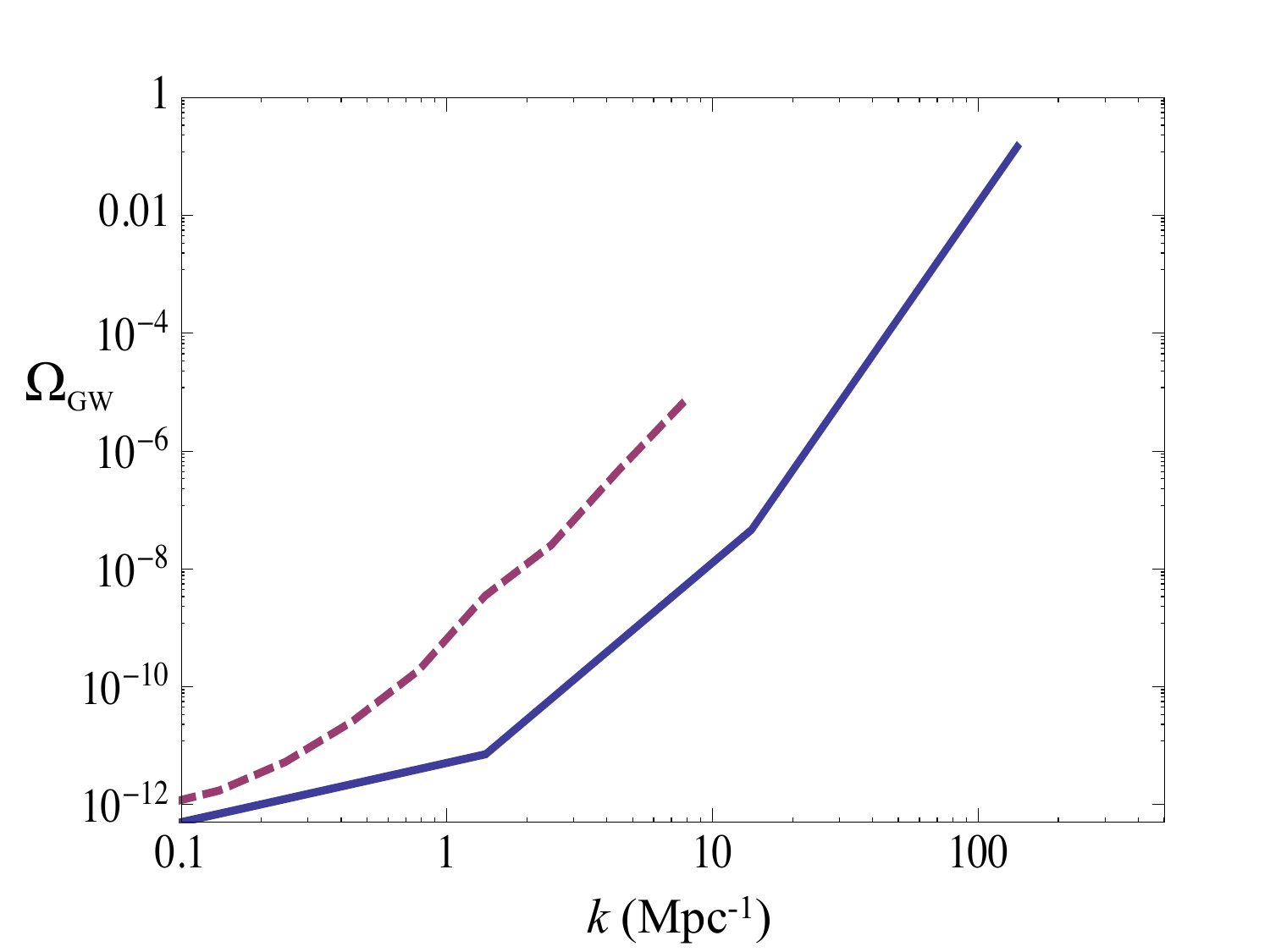}
\caption{Solid line is the upper bound on the spectral energy density of the SGWB derived from Planck, ACT, and SPT small-scale CMB temperature anisotropy measurements in this work. The dashed line, drawn from the blue solid curve in Figure 4 of Ref.~\cite{namikawa}, is the upper bound derived from a likelihood analysis using the CMB temperature anisotropy data made by Planck.}
\label{fig2}
\end{figure}

\subsection{CMB constraints on SGWB}

In Ref.~\cite{namikawa}, the authors have performed a likelihood analysis using the CMB anisotropy and polarization data from Planck and BICEP/Keck array to derive upper bounds on the SGWB for $0.1\,{\rm Mpc}^{-1}\lesssim k \lesssim 10\,{\rm Mpc}^{-1}$. In their results, they have placed an upper limit on the contribution of tensor modes to the primary CMB temperature anisotropy for $l\le 2500$, denoted by the dashed line in Fig.~\ref{fig2}.
In the present work, we will simply use the CMB anisotropy power spectra induced by SGWB 
in Fig.~\ref{fig1} to set bounds on the SGWB. 

Combining Eqs.~(\ref{omegaGW2}) and (\ref{Cl4narrow}), we obtain
\begin{equation}
{\cal D}_l\equiv T_0^2 l(l+1)C_l/(2\pi)=
\frac{3H_0^2}{2k_*^2}\Omega_{\rm GW} T_0^2 D_l,
\end{equation}
where $T_0=2.725{\rm K}$ is the present CMB temperature and $D_l$ is defined in Fig.~\ref{fig1}.
The measured primary CMB anisotropy power spectrum at $l=1900$ is given by $200\mu{\rm K}^2$~\cite{planck19,act20,spt21}. The statistical detection of the secondary CMB anisotropies at $l=3000-10^4$ made by both the Atacama Cosmology Telescope (ACT) and the South Pole Telescope (SPT) is at a level of $3\mu{\rm K}^2$~\cite{act20,spt21},
which is an inferred value of the secondary CMB temperature anisotropy based on the model involving various contributors and the foreground removal scheme.

In Fig.~\ref{fig1}, we have $D_{1900}\simeq 10^7$ for $k =0.14\,{\rm Mpc}^{-1}$. Requiring that this anisotropy power is less than the measured value, i.e. ${\cal D}_{1900}\lesssim 200\mu{\rm K}^2$, we obtain
$\Omega_{\rm GW} h^2 \lesssim 3\times 10^{-13}$.
For $1.4\,{\rm Mpc}^{-1}\le k  \le140\,{\rm Mpc}^{-1}$, we read the three power spectra $D_{10^4}$'s for $k=1.4$, $14$, and $140\,{\rm Mpc}^{-1}$ from Fig.~\ref{fig1}. Assuming that each $D_{10^4}$ cannot exceed the inferred
value of the secondary CMB contribution, i.e. ${\cal D}_{10^4}\lesssim 3\mu{\rm K}^2$, we set upper limits on 
$\Omega_{\rm GW} h^2$ at $k=1.4$, $14$, and $140\,{\rm Mpc}^{-1}$. 
Then, we interpolate linearly between these four single-point upper limits. The resultant upper bound is given
by the solid line in Fig.~\ref{fig2}, where $h=0.67$ is assumed.

In Fig.~\ref{fig2}, the value of the upper bound (solid line) in this work at $k =0.1\,{\rm Mpc}^{-1}$ is about equal to that (dashed line) obtained in Ref.~\cite{namikawa}. This would be the case because both values are derived by using the Planck measured primary CMB anisotropy power spectrum. 
For $1\,{\rm Mpc}^{-1}\lesssim k  \lesssim 10\,{\rm Mpc}^{-1}$, using the inferred value of the secondary CMB anisotropy at $l=10^4$ by ACT and SPT, we have obtained more stringent limits than those obtained from the Planck data in Ref.~\cite{namikawa}. Furthermore, we have extended the range for the upper bound to $k \le 140\,{\rm Mpc}^{-1}$.

\section{Pulsar timing}
\label{ptiming}

In the current pulsar-timing observation, radio pulses from an array of roughly 100 Galactic millisecond pulsars are being monitored with ground-based radio telescopes~\cite{romano}. The redshift fluctuation of a pulsar in the pointing direction ${\bf e}$ on the sky is given by
\begin{equation}
z({\bf e})= - {1\over 2}\int_{\eta_e}^{\eta_r}d\eta\, e^ie^j\frac{\partial}{\partial\eta}h_{ij}(\eta, \vec x),
\label{rsfze}
\end{equation}
where we have used ${\bar z}=0$ in Eq.~(\ref{Deltaz}) since the pulsar is in our Galaxy. 
The physical distance of the pulsar from us is $D=\eta_r-\eta_e$, which is of order $1\,{\rm kpc}$.
The quantity that is actually observed in the pulsar-timing observation is the time residual counted as
\begin{equation}
r(t)=\int_0^t dt'z(t'),
\end{equation}
where $t'$ denotes the laboratory time and $t$ is the duration of the observation. 
Using the laboratory time $t'$, we rewrite Eq.~(\ref{rsfze}) as
\begin{equation}
z(t',{\bf e})= - {1\over 2}\int_{t'+\eta_e}^{t'+\eta_r}d\eta\,e^ie^j\frac{\partial}{\partial\eta}h_{ij}(\eta, \vec x).
\end{equation}

Let us consider a SGWB with the narrow power spectrum~(\ref{Npower}), 
where $M_*/M_p$ is the present GW amplitude. The wavenumber is assumed to be $k_* \sim 10^6 {\rm Mpc}^{-1}$,
lying within the pulsar-timing-array sensitivities to GWs at nanohertz frequencies.
At the present time, the GWs are traveling plane waves:
\begin{equation}
h(k_*\eta)\simeq  \frac{M_*}{M_p} e^{-ik_*\eta}.
\end{equation}
Then, we can construct the time-residual correlation between a pair of pulsars:
\begin{eqnarray}
&&\langle r(t_1)r(t_2) \rangle = \int_0^{t_1}dt' \int_0^{t_2}dt{''} \langle z(t') z(t'') \rangle \nonumber \\
&&= \int_0^{t_1}dt' \int_0^{t_2}dt{''}  e^{-ik_*(t'-t'')} \langle z({\bf e_1}) z({\bf e_2}) \rangle.
\label{rcorr}
\end{eqnarray}
In Eq.~(\ref{rcorr}), the integrand is simply the redshift fluctuation correlation, 
\begin{equation}
\langle z({\bf e_1}) z({\bf e_2}) \rangle=\sum_l \frac{2l+1}{4\pi} C_l P_l({\bf e}_1\cdot{\bf e}_2),
\label{zcorr}
\end{equation}
whose power spectrum is given by
\begin{equation}
C_l\simeq\frac{M_*^2}{2\pi M_p^2} (l+2)(l+1)l(l-1) \left\vert\int_{0}^{x_1} dx\, e^{ix}\frac{j_l(x)}{x^2}\right\vert^2,
\label{galacticCl}
\end{equation}
where $x=k_*(\eta_r-\eta)$ and $x_1=k_*D\sim 10^3$. 
Using the approximation  in Eq.~(\ref{besselintegral}), we obtain an exact form for the power spectrum as
\begin{equation}
C_l\simeq\frac{2M_*^2}{\pi M_p^2} [(l+2)(l+1)l(l-1)]^{-1}.
\label{scalingCl}
\end{equation}
In fact, this $l^{-4}$ scaling has been derived using other methods~\cite{gair}. In Ref.~\cite{gair}, the authors have also shown that substituting the power spectrum~(\ref{scalingCl}) into the two-point correlation function~(\ref{zcorr}) would give us the Hellings and Downs curve for the quadrupolar interpulsar correlations~\cite{downs},
which is given by the earth term in the Shapiro time delay of the arrival times of radio pulses from pulsars.

Recently, the NANOGrav Collaboration~\cite{nanograv} has found strong evidence of a stochastic common-spectrum process 
across 45 millisecond pulsars, alluding to a SGWB with a characteristic strain of $h_c=1.92\times10^{-15}$ 
at a reference frequency of $f_{\rm yr}=1\,{\rm yr^{-1}}\simeq 31.8\,{\rm nHz}$. 
However, they have not found statistically significant evidence that this process 
has Hellings and Downs spatial correlations. 
If the SGWB is confirmed, its spectral energy density at $f_{\rm yr}$ can be read from Eq.~(\ref{omegaGW2}) as
\begin{eqnarray}
\Omega_{\rm GW} h^2 &\simeq&  \frac{1}{6\pi^2} \frac{M_*^2}{M_p^2} \left(\frac{k_*}{100\,{\rm km\,s^{-1}Mpc^{-1}}}\right)^2 \nonumber \\
&=&\frac{2\pi^2}{3} h_c^2 \left(\frac{f_{\rm yr}}{100\,{\rm km\,s^{-1}Mpc^{-1}}}\right)^2 \nonumber \\
&\simeq& 2.3 \times 10^{-9},
\end{eqnarray}
where $|h(k_*\eta)|=M_*/M_p=\pi h_c$ and $k_*=2\pi f_{\rm yr}\simeq 2.06\times 10^7 {\rm Mpc}^{-1}$.

\begin{figure}[htp]
\centering
\includegraphics[width=0.5\textwidth]{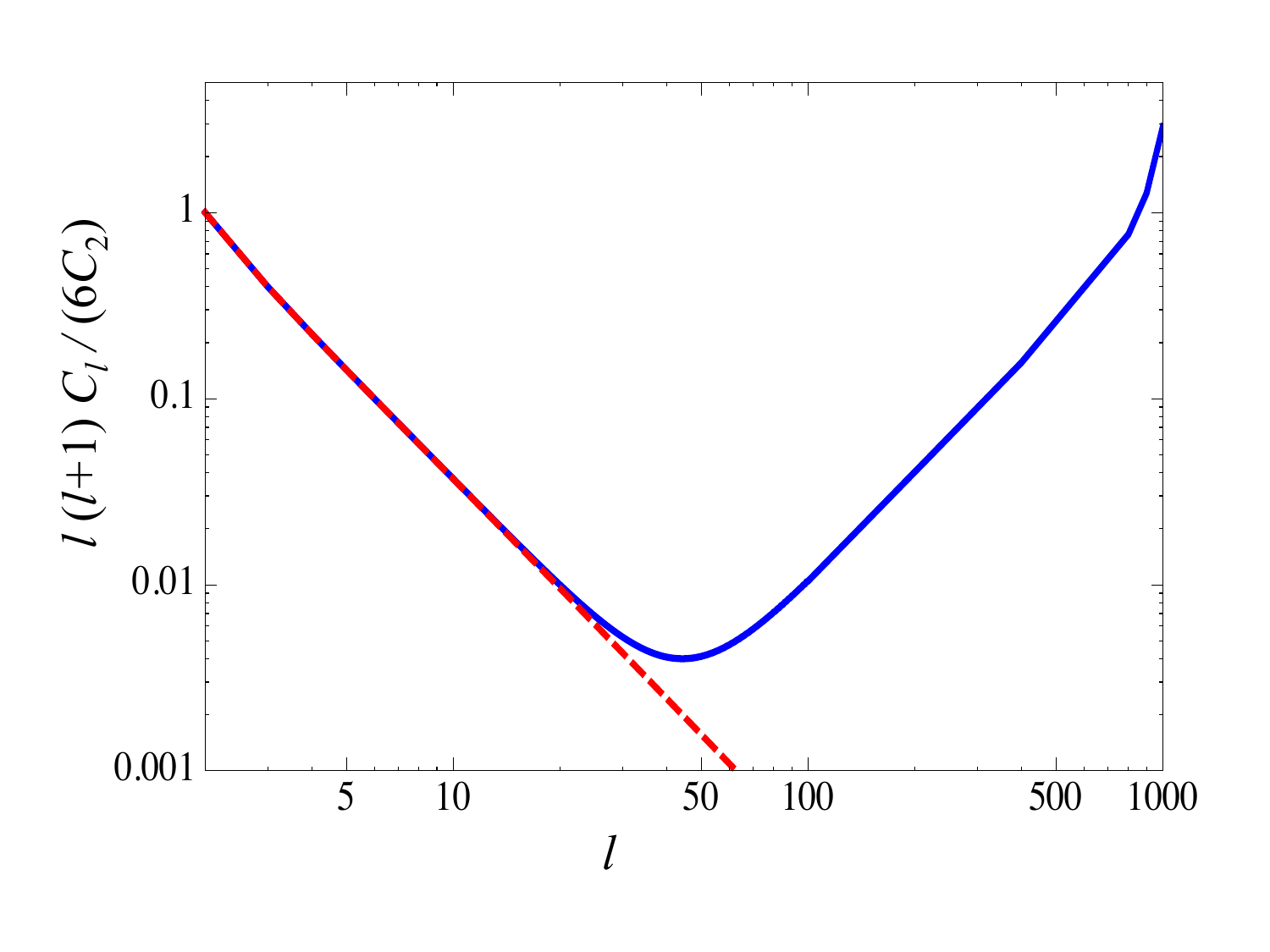}
\caption{Power spectra of the redshift fluctuation correlation. The dashed line is the Hellings and Downs power spectrum.
The solid line is drawn from computing numerically Eq.~(\ref{galacticCl}) for $x_1=k_* D=10^3$.}
\label{fig3}
\end{figure}

However, the exact form~(\ref{scalingCl}) underestimates the values of $C_l$ at large $l$'s.
In Fig.~\ref{fig3}, we have plotted the power $l(l+1)C_l$ against $l$, using the true value $x_1=10^3$ to numerically evaluate the integral in Eq.~(\ref{galacticCl}). 
The resulting power spectrum is close to the Hellings and Downs spectrum~(\ref{scalingCl}) on large angular 
scales when $l<20$, with $C_l$ increased by $0.002\%$, $0.3\%$, $4.4\%$, $21.6\%$, $67.2\%$ at $l=2,10,20,30,40$, respectively. There exists a significant power at small scales when $l\lesssim 10^3$. 
When we observe the pulsars at distance $D$, the angular separation between them for us to see the spatial fluctuation of GWs with wavelength $\lambda$ is $\lambda/D$. This explains why $C_l$ increases at small scales and peaks 
at $l\sim 10^3$. 
This small-scale power does not change the Hellings and Downs curve on large angular scales, 
while giving a sharp peak to the curve at small separation angles~\cite{chu2107}. 
In Fig.~\ref{fig3}, the power spectrum is roughly a v-shape line standing at $\log l \sim 1.5$ (or $l\sim 40$), 
which separates between the large-scale power and the small-scale power.
This is anticipated from the fact that the autocorrection has a power twice larger than the Hellings and Downs curve at zero lag (see, for example, Ref.~\cite{gair}) induced by the pulsar term of the Shapiro time delay.
As such, it would be interesting to search for this small-scale power by measuring correlation 
between adjacent pulsars separated by about $180^\circ/l \sim 0.2^\circ$ ($l\sim 10^3$) on the sky.
For nearby pulsars with $D\sim 0.1\,{\rm kpc}$, the exact form in Eq.~(\ref{scalingCl}) is no longer a good approximation, 
so one should use the full Eq.~(\ref{galacticCl}) to compute the power spectrum.

The integral in Eq.~(\ref{galacticCl}) is evaluated assuming that all the pulsars are at the same distance.
However, in realistic observation they are spread out in distance. As such, the coherence will be lost, 
resulting in a suppression of the small-scale power. To assess the loss of coherence, let us consider a pair of pulsars with a sub-degree angular separation in a globular star cluster at distance of $1\,{\rm kpc}$, noting that the size of a globular cluster ranges from a few pc to less than $0.1\,{\rm kpc}$. Suppose one of the pulsar pair is nearer to us than the other one by 
$\Delta x_1$; then, from Eq.~(\ref{galacticCl}) the fractional change in $C_l$ will be given by
\begin{equation}
\frac{\Delta C_l}{C_l} = - \left[\int_{x_1-\Delta x_1}^{x_1} dx\, e^{ix}\frac{j_l(x)}{x^2}\right]
\left[\int_{0}^{x_1} dx\, e^{ix}\frac{j_l(x)}{x^2}\right]^{-1}\,.
\end{equation}
When $x_1=10^3$ and $\Delta x_1=1$ (giving $\Delta D=1\,{\rm pc}$), $\vert\Delta C_l/C_l\vert <1 $ for $l<10^3$, so the small-scale power still remains. When $\Delta D$ increases to $10\,{\rm pc}$, $\vert\Delta C_l/C_l\vert <1 $ as long as $l<400$. A search for this small-scale power in the current pulsar-timing observation is difficult
due to poor statistics from a limited number of monitored pulsars on the sky.
The future SKA project will observe about 6000 Galactic millisecond pulsars to reach a sensitivity three to four orders of magnitude better than the current pulsar-timing-array experiments~\cite{SKA}. It would be interesting to hunt for pulsar pairs in globular clusters to measure the correlation at small angular scales.

Furthermore, it would be interesting to consider extragalactic millisecond pulsars or other presumable cosmological precision clocks to measure the SGWB. 
In this case, $\eta_r=\eta_0$ and $\eta_e$ is the time of emission of light from the
extragalactic sources at redshift ${\bar z}$. Assume $\eta_e>\eta_*$. Then, the redshift-fluctuation correlation function
is enhanced by the redshift factor and reads
\begin{equation} 
\langle \delta z({\bf e_1}) \delta z({\bf e_2}) \rangle= (1+{\bar z})^2 \sum_l \frac{2l+1}{4\pi} C_l P_l({\bf e}_1\cdot{\bf e}_2).
\end{equation}
Here $C_l$ is given by Eq.~(\ref{extraCl}) with $x_1=k_*D\gg 10^3$, 
where $D=\eta_0-\eta_e$ is the comoving distance to the extragalactic sources.

\section{Conclusion}
\label{conclusion}

We have revisited the Sachs-Wolfe gravitational effect of the stochastic gravitational wave background.
Considering the effect as redshift-space fluctuations integrated along the line-of-sight from the observer
to the observable, we have found that the line-of-sight integral is particularly useful for studying the 
imprint of short-wavelength gravitational waves on the CMB anisotropy, without having recourse to 
intensive numerical computations. The integral in Eq.~(\ref{Cl4narrow}) is the main result for us to compute
the CMB anisotropy power spectra induced by short-wavelength SGWB. Thus,
we have found that the contribution of short-wavelength gravitational waves
to the large-scale CMB anisotropy $C_l$ scales as $l^{-4}$. Furthermore, we have derived the new constraints 
on the SGWB using Planck, ACT, and SPT small-scale CMB anisotropy data.

The Sachs-Wolfe gravitational effect can well be used to study the redshift fluctuations of millisecond pulsars. 
The time-residual correlation between a pair of pulsars can then be expressed in terms of a power spectrum given by
the exact line-of-sight integral in Eq.~(\ref{scalingCl}). This reproduces the Hellings and Downs curve for 
the redshift correlation between a pair of distant and separated pulsars. For nearby pulsars or close pulsar pairs, 
we have calculated the deviations from the Hellings and Downs curve that should be taken into account in
pulsar timing measurements, in particular when the correlation on small angular scales comes into an important role.
Our results will be useful for future pulsar-timing arrays that observe thousands of millisecond pulsars.

\begin{acknowledgments}
This work was supported in part by the Ministry of Science and Technology (MOST) of Taiwan, Republic of China, under
Grant No. MOST 109-2112-M-001-003.
\end{acknowledgments}

\raggedright

\end{document}